\documentclass[aps,prb,twocolumn,superscriptaddress]{revtex4-2}
\usepackage{hyperref}
\usepackage[utf8]{inputenc}
\usepackage[english]{babel}
\usepackage{amsmath}	
\usepackage{bm}
\usepackage{amsfonts}
\usepackage{mathtools}
\usepackage{relsize}
\usepackage{threeparttable}
\usepackage{amssymb}
\usepackage{xcolor}
\usepackage{enumerate}
\usepackage{verbatim}
\usepackage{mathrsfs}
\usepackage{caption}
\usepackage{subcaption}

\usepackage{braket}
\usepackage{xr}
\usepackage{graphicx}
\usepackage[left=2cm,right=2cm,top=3cm,bottom=3cm]{geometry}

\newcommand{\R}{{\bf r}}
\newcommand{\K}{{\bf k}}
\newcommand{\RR}{{\bf R}}
\newcommand{\abs}[1]{\left\lvert{#1}\right\rvert}

\begin{document}

\title{Real-space understanding of electron-phonon coupling in superconducting hydrides}

\author{Trinidad Novoa}
    \email[]{trinidadantonia.novoa@ehu.eus}
    \affiliation{Fisika Aplikatua Saila, Gipuzkoako Ingeniaritza Eskola, University of the Basque Country (UPV/EHU), Europa Plaza 1, 20018 Donostia/San Sebastián, Spain}
    \affiliation{Centro de Física de Materiales (CSIC-UPV/EHU), Manuel de Lardizabal Pasealekua 5, 20018 Donostia/San Sebastián, Spain }
\author{Francesco Belli}
    \affiliation{Department of Chemistry, State University of New York at Buffalo, Buffalo, NY 14260-3000, USA}
\author{Raffaello Bianco}
    \affiliation{Dipartimento di Scienze Fisiche, Informatiche e Matematiche, Università degli Studi di Modena e Reggio Emilia, Modena, Italy}
    \affiliation{Centro S3, Istituto Nanoscienze-CNR, Modena, Italy}
\author{Julia Contreras-Garc\'ia}
    \affiliation{Laboratoire de Chimie Th\'eorique, Sorbonne Universit\'e \& CNRS, 4 Pl. Jussieu, 75005, Paris, France }
\author{Ion Errea}
    \email[]{ion.errea@ehu.eus}
    \affiliation{Fisika Aplikatua Saila, Gipuzkoako Ingeniaritza Eskola, University of the Basque Country (UPV/EHU), Europa Plaza 1, 20018 Donostia/San Sebastián, Spain}
    \affiliation{Centro de Física de Materiales (CSIC-UPV/EHU), Manuel de Lardizabal Pasealekua 5, 20018 Donostia/San Sebastián, Spain }
    \affiliation{Donostia International Physics Center (DIPC), Manuel de Lardizabal Pasealekua 4, 20018 Donostia/San Sebastián, Spain }

\date{\today}

\begin{abstract}
Electron-phonon coupling is at the origin of conventional superconductivity, enabling the pairing of electrons into Cooper pairs. The electron-phonon matrix elements depend on the electronic eigenstates and, in the standard linear approximation, on the first derivative of the potential felt by the electrons with respect to ionic perturbations. Here, we focus on the derivatives of the potential with a twofold aim: to assess their contribution to the overall coupling and to analyze the limitations of neglecting higher-order derivatives. Several real-space functions are proposed to do the analysis, and are computed for some well-known superconductors. Our results show that, in hydrides, the derivatives of the potential tend to be larger in regions of high electron localization, explaining the success of electronic descriptors previously observed to correlate with the critical temperature. The new functions introduced here are able to tell apart structures with similar types of bonding but very different critical temperatures, such as H$_3$S and H$_3$Se $Im\bar{3}m$ phases, where electronic descriptors alone fail. Moreover, they provide a method to discriminate promising superconductors from conventional low-$T_c$ materials. Interestingly, our descriptors are capable of easily estimating the impact of higher-order terms in the electron-phonon coupling. In fact, we capture the limitations of the linear approximation expected for PdH, and predict an even more important non-linear behavior in other hydrides. 
\end{abstract}

\maketitle

\section{Introduction}

Over the last century, scientists have been searching for superconductivity at ambient conditions. During this time, 
two very different families of superconductors have come closer to reaching room critical temperatures: cuprates and hydrogen-rich compounds. The former pose the great disadvantage of being unconventional, meaning that no theory accounts for superconductivity in those materials. 
Meanwhile, superconducting hydrides are stable only at megabar pressures \cite{Drozdov2015,Drozdov2019,Kong2021,chen2021high,Semenok2021,bi2022,Li2022natcomm,song2023,Saha2023,chen2023,pinsook2025}, 
even if practically room temperature has been achieved for them. Nonetheless, they remain a topic of intense research given the prospects of discovering high-critical temperature ($T_c$) hydrides at low or even ambient pressures~\cite{gao2025,dangic2024-rbph3,sanna2024,Cerqueira2024,Dolui2024Feasible,FLORESLIVAS20,pickard2020,Hutcheon2020,hilleke2022,wei2023,dicataldo2023,fang2024,zheng2024} and the conventional nature of their superconductivity, which allows the \emph{ab initio} estimation of their $T_c$ through the calculation of the electron-phonon interaction. 

Theoretical estimations of $T_c$ using the Migdal-Eliashberg formalism require calculating the electronic wavefunctions and the first order derivatives of the effective electronic potential \cite{Migdal1958,Eliashberg1960,Marsiglio2020}. This, combined with the vastness of the chemical space of hydrides, makes theoretical prediction of new materials a cumbersome task.
Using machine learning to accelerate the discovery of new superconductors is thus an attractive alternative, for which more than one approach has been adopted recently. On the one hand, it is possible to train the critical temperature directly from the chemical composition and structural information \cite{stanev2018}, or to add on top cheap electronic quantities like the density of states to improve the predictions of $T_c$ \cite{Cerqueira2024, cerqueira2024-2}. This kind of models allow to make very fast predictions, but training them is costly and requires extensive databases. Another approach is to train $T_c$ using higher-level input quantities such as the Eliashberg function \cite{xie2022}, which yields more accurate results but does not greatly reduce computation times with respect to standard first-principles calculations based on Migdal-Eliashberg theory. Finally, one can define cheap descriptors inspired by concepts used in quantum chemistry to characterize materials, and use them to train $T_c$~\cite{Belli21,dimauro2024,belli2025}. This option not only accelerates and improves the training, but it also provides physical insight into the properties that characterize high-$T_c$ superconductors, as they contain more information on the electronic properties of the material, and describe the superconducting state \cite{muriel2024}.
In particular, in Refs.~\citenum{Belli21} and \citenum{dimauro2024}, electronic descriptors derived from the density of states (DOS) at the Fermi energy and from the electron localization function (ELF)\cite{Becke1990,SavinELF} were proposed. Both of them can be easily obtained in standard packages performing density functional theory (DFT) calculations, and the estimation of $T_c$ can be obtained as a post-processing procedure that only takes a couple of seconds \cite{novoa2024}. Nonetheless, those estimations are still far from accurate, which motivates the ongoing search for descriptors that could improve the characterization of hydride superconductors.

The ability of electronic properties to describe and predict superconductivity in conventional materials has been described in the literature several times\cite{Belli21, dimauro2024,Li2022, lavroff2024, yu2024,belli2025}. However, because we are dealing with electron-phonon superconductors, we expect that the inclusion of vibrational properties should improve that description.
In this context, the question arises as to whether the other part of the electron-phonon coupling -- the one accounting for the changes of the ionic potential due to the vibrations -- plays a significant role. 
Despite the previous observations suggesting that its role 
could be less important, a thorough study to confirm or deny that hypothesis is lacking. Moreover, because said correlations have only been proposed for hydrogen-based compounds, it is also in our interest to study the differences between conventional superconductors with and without hydrogen. We expect those differences to be important, owing to two very particular electronic and vibrational properties of hydrogen: that it does not have a core, and that the light weight of its nuclei can lead to large quantum ionic fluctuations.

In fact, in hydrides the phonon modes become highly anharmonic as a consequence of those quantum fluctuations, giving rise to non-negligible effects on the materials' structural and dynamical properties \cite{errea2016,errea2020,hou2021,belli2022}. The breakdown of the harmonic approximation is a reflection of the insufficiency of the second-order expansion of the Born-Oppenheimer (BO) potential to properly describe the ionic vibrations, which is a direct consequence of the large displacements of the hydrogen ions. In the case of the BO potential, a more adequate representation can be obtained by including higher-order derivatives in a non-perturbative way, for instance, within the stochastic self-consistent harmonic approximation (SSCHA)~\cite{errea2014,monacelli2021}. The electron-phonon coupling is usually also calculated with a lowest-order linear expansion of the Kohn-Sham potential with respect to ionic displacements.
However, considering the magnitude of the ionic displacements in hydrides, it seems contradictory to expect that only the linear part of the expansion should be enough for the electron-phonon coupling, but that anharmonic effects in the BO potential need to be considered for the structural and dynamical properties.

Estimating the electron-phonon coupling beyond the linear approximation is very cumbersome and few materials have been studied thus far from first principles~\cite{heid1992,bianco2023,Liu2020,Chen2021,Chen2022,Zhang2022,Chen2024}.
In Ref.~\citenum{bianco2023}, this was done for palladium hydride and  aluminum, confirming that  non-linear effects are significantly more pronounced in the superconductor containing hydrogen. For PdH, a linear approach led to large discrepancies in the electron-phonon coupling constant, which is crucial for estimating $T_c$. 
Little is known about similar behaviors in other hydrides, and the consequences that this could have on the estimation of electron-phonon coupling and critical temperature in those materials. Although vertex corrections to electron-phonon coupling have proven to be non-negligible in H$_3$S \cite{mishra2025,mishra2026}, they have only been studied in a perturbative fashion, without capturing the non-linear effects of the KS potential.

By including explicitly the analysis of the derivatives of the Kohn-Sham potential, in this paper we present new real-space functions and descriptors to predict $T_c$ that improve previous ones in the literature only based on electronic properties. We also demonstrate how these can be used to anticipate the role of non-linear effects on the electron-phonon coupling, providing new insight into these barely studied effects. 
This paper is structured as follows. In Section \ref{sec:e-ph-coupling}, we introduce electron-phonon coupling and show how it can be written in terms of electronic descriptors and derivatives of the electronic effective potential. In Section \ref{sec:real-space}, we propose four different real-space descriptors and analyze their magnitudes in a group of well-known conventional superconducting materials. The purpose of the analysis is twofold: (i) to assess the contribution of the derivatives of the potential to the electron-phonon coupling in hydrides, proposing a way of including those in a future model to predict $T_c$ with more accuracy; and (ii) to evaluate the importance of non-linear effects of the potential in hydride superconductors. Finally, in Section \ref{sec:conclusions} we present a summary of the results and some conclusions.

\section{Electron-phonon coupling}
\label{sec:e-ph-coupling}

Under the Born-Oppenheimer (BO) approximation, ionic and electronic degrees of freedom are decoupled, and one can solve the system as two separate problems. Normally, the electronic Hamiltonian, $\mathcal{H}_e$, considers that the $M$ nuclei and its inner cores are fixed in some ionic positions, $\RR=\{\RR_1,\cdots,\RR_M\}$, that are treated parametrically. In density functional theory, it is possible to find the one-electron wavefunctions, $\psi_i(\R; \RR)$, by solving the Kohn-Sham (KS) equations,
\begin{align}
    \left[-\frac{1}{2}\nabla_{\R}^2 + v_{KS}(\R; \RR) \right] \psi_i(\R; \RR) = \varepsilon_i \psi_i(\R; \RR)\, ,
\end{align}
where $v_{KS}(\R;\RR)$, the Kohn-Sham potential, is a one-body effective potential acting on the electrons. It contains the attractive ionic potential, the Coulomb repulsion between electrons, and an approximate term accounting for exchange and correlation. 

In the case of a crystal, we consider a supercell containing $N_c$ unit cells, satisfying periodic boundary conditions (PBCs), where eigenstates are the Bloch functions, $\psi_i(\R)= \psi_{n\K}(\R) = \braket{\R |n\K}$. The indices $n,\K$ indicate both the band index and the momentum, respectively. When ions are perturbed out of their equilibrium positions, 
the KS potential felt by the electrons also fluctuates. It is thus possible to measure the coupling between the electrons and the phonons in a crystal by assessing the magnitude of those changes. In fact, the electron-phonon matrix elements depend on the derivatives of $v_{KS}$ with respect to lattice distortions,
\begin{align}
g^a_{m\K, n\K'} = \left\langle m\K\left\lvert \left.\frac{\partial v_{KS}(\R;\RR)}{\partial u_a}\right\rvert_{u_a=0} \right\rvert n\mathbf{k}'\right\rangle\, ,
\label{eq:e-ph-matrix}
\end{align}
where $u_a$ are the displacements of the ionic positions from equilibrium, and the subscript $a=(i\alpha)$ refers to the ionic site $i$ in the supercell, and the Cartesian coordinate $\alpha$. For simplicity, we consider distortions from equilibrium that have the periodicity of the supercell and are otherwise completely general, i.e.~we do not consider monochromatic perturbations with finite wavevector $\mathbf{q}$, that would imply distortions not conmensurate with the supercell. The electron-phonon coupling constant,
\begin{align}
    \lambda = 2\int d\omega \frac{\alpha^2 F(\omega)}{\omega}\, ,
    \label{eq:e-ph-lambda}
\end{align}
is often used to assess the magnitude of the coupling, and depends directly on the matrix elements through the Eliashberg function,
\begin{align}
    \alpha^2 F(\omega) &= \sum_{a,b} \Delta^{ab} B_{ab} (\omega) \, ,
    \label{eq:e-ph-eliash}
\end{align}
where $\Delta_{ab}$ is the averaged deformation potential over the Fermi surface, 
\begin{align}
\notag
    \Delta_{ab} = \frac{1}{N_FN_{\K}} \sum_{n,m,\K,\K'} &g^a_{m\K,n\K'}\, g^b_{n\K',m\K} \, \times \\
    &\times \delta(\varepsilon_{m\K}-\varepsilon_F)\delta(\varepsilon_{n\K'} - \varepsilon_F)\, ,
    \label{eq:e-ph-delta}
\end{align}
and $B_{ab}(\omega)$ is the phonon spectral function.\cite{allen1983,dangic2024}
In Eq. \eqref{eq:e-ph-delta} $N_F$ is the density of states per spin at the Fermi level and $N_{\K}$ the number of $\K$ points in the sum.

\section{Real-space descriptors of the derivatives of the KS potential}
\label{sec:real-space}

Equation \eqref{eq:e-ph-matrix} puts into evidence that the coupling between electrons and phonons depends on two things: the derivatives of the KS potential with respect to ionic displacements, and the overlap between the electronic states. Separately analyzing those contributions should provide valuable insight into the key factors that determine superconductivity.
The contribution of the derivatives of $v_{KS}$ on the magnitude of electron-phonon coupling and critical temperature of hydrogen-based superconductors has not been studied before in the literature. After reviewing the electronic descriptors that allow to describe hydride superconductors,
the aim of this section is to analyze, from first principles, the real-space derivatives of the KS potential and its relationship to the critical temperature of some well-known superhydrides.

\subsection{Electronic descriptors}

The utility of electronic real-space descriptors to characterize hydrogen-based superconductors has now been demonstrated many times
, with the electron localization function (ELF) being particularly good at describing those materials. It is defined as
\begin{align}
    ELF(\R) = \left[1+ \left(\frac{\tau(\R) - \tau_{vW}(\R)}{\tau_{TF}(\R)}\right)^2 \right]^{-1}\, ,
\end{align}
with $\tau(\R)$ the kinetic energy density of the system, depending on the KS wavefunctions, while $\tau_{vW}(\R)$ and $\tau_{TF}(\R)$ are the von Weizsäcker and Thomas-Fermi approximations to it, which depend directly on the electron density \cite{Becke1990,SavinELF,pendas2023}.

The maxima of the ELF are located in regions where opposite-spin electrons are likely to be found, providing a quantitative way of assessing the types of bond present in the lattice, which in turn allows a categorization of hydrogen-rich compounds into different families. This permits to identify high-$T_c$ compounds, which are usually either of covalent character, or present a clathrate-like structure with hydrogens weakly bonded to each other in elongated bonds \cite{Belli21}. The molecularity index ($\phi^*$), defined as the highest ELF value for which two hydrogens connect within the same isosurface,  automatizes the bond-type categorization and therefore the screening of promising compounds \cite{dimauro2024}.

Perhaps the most useful electronic descriptor proposed so far is the networking value, $\phi$, introduced by Belli \emph{et al.}~in Ref.~\citenum{Belli21}, defined as the highest ELF isovalue that creates a fully connected isosurface that is periodic in three dimensions. This quantity measures the degree of electronic delocalization, and it shows a positive correlation with $T_c$, which can be improved by including some extra ingredients: the fraction of hydrogen atoms in the unit cell, $H_f$, and the fraction of the density of states at the Fermi energy that projects onto hydrogen orbitals, $H_{DOS}$. 

Several fits have been proposed to estimate $T_c$ from those four descriptors ($\phi$, $\phi^*$, $H_f$, and $H_{DOS}$) \cite{Belli21, dimauro2024, novoa2024,belli2025}. In the following, we use the Gradient Boosting Regression (GBR) fit proposed in Ref.~\citenum{novoa2024}, as a reference for predictions of $T_c$ that consider purely electronic information. This is done using TcESTIME, a program that automatically searches for the ELF-derived descriptors and consequently estimates $T_c$ \cite{novoa2024, TcESTIME}.

All in all, the electronic descriptors have boosted a better understanding of what drives superconductivity in hydrides from a chemical point of view. The estimations of $T_c$ are useful because of their simplicity, but they lack accuracy, with predictions showing mean absolute errors around 33K in $T_c$, and even rising to 100 K in a few problematic compounds.\cite{novoa2024} This motivates the search for other descriptors that include the effect of the derivative of the KS potential due to ionic vibrations on the electron-phonon coupling. 

\begin{table*}[htbp]
    \centering
    \caption{Summary of reference values and results for hydrides, including: pressure (P), networking value ($\phi$), molecularity index ($\phi^*$), DOS hydrogen fraction ($H_{DOS}$), hydrogen fraction ($H_f$), $D^2$,  predicted $T_c$ using electronic descriptors obtained with GBR fit of TcESTIME ($T_c^{\rm GBR}$), and reference critical temperature obtained from the literature ($T_c^{\rm ref}$), as well as the work from where those were taken.}
    \label{tab:table_all-1}
    \begin{tabular}{cccccc|c|ccc}
    \hline
        & P (GPa) & $\phi$ & $\phi^*$ & $H_{DOS}$  & $H_f$ & $D^2(Ry^2/a_0^2)$& $T_c^{\rm GBR}$ (K) &  $T_c^{\rm ref}$ (K) & Ref.\\
        \hline \hline
        PdH        & 0   & 0.21 & 0.21 & 0.101 & 0.500 & 0.49 &   9 & 5   & \cite{errea2013}\\
        H$_3$Se    & 200 & 0.66 & 0.66 & 0.479 & 0.750 & 1.58 & 194 & 116 & \cite{zhang2015} \\
        H$_3$S     & 200 & 0.69 & 0.69 & 0.473 & 0.750 & 1.94 & 201 & 204 & \cite{duan2014} \\
        H$_3$S     & 150 & 0.71 & 0.71 & 0.488 & 0.750 & 1.82 & 223 & 225 & \cite{errea2016}\\
        LaH$_{10}$ & 150 & 0.56 & 0.67 & 0.591 & 0.909 & 1.73 & 253 & 250 & \cite{errea2020}\\        
        LaBH$_{8}$ & 100 & 0.32 & 0.60 & 0.626 & 0.800 & 1.67 & 87 & 137 & \cite{belli2022}\\
        \hline \hline
    \end{tabular}   
\end{table*}

\subsection{Systems of study}

In this work, we analyze the derivatives of the Kohn-Sham potential in real space in some well-known superconductors. Two compounds with high critical temperature but different chemical bonding are studied, namely the clathrate-like $Fm\bar{3}m$-LaH$_{10}$ structure at 150 GPa, with calculated $T_c=250$ K,\cite{Drozdov2019,errea2020} and the covalent $Im\bar{3}m$-H$_3$S structure at 150 and 200 GPa, with $T_c$'s of 225 K and 204 K, respectively, also theoretically estimated.\cite{Drozdov2015,errea2016,duan2014} 
A similar covalent structure where sulfur is replaced by selenium is included in the analysis, which is predicted to have a considerably lower critical temperature of 116 K at 200 GPa,\cite{zhang2015} nearly half of that of H$_3$S. This is a case where the electronic descriptors fail to capture the differences between the two systems (see Table~\ref{tab:table_all-1}), which share similar ELF patterns and have close networking values of 0.66 and 0.69 for H$_3$Se and H$_3$S, respectively. 
Palladium hydride is also included in the analysis, a compound without any  covalent bonds and where Pd and H atoms remain isolated, for which $T_c=5$ K according to theoretical predictions.\cite{errea2013} Finally, the ternary hydride LaBH$_8$ in its $Fm\bar{3}m$ phase at 100 GPa is added to the set, with a reference $T_c$ of 137 K \cite{belli2022}. For comparison, 
we include the aluminum crystal, 
a conventional superconductor that does not contain hydrogen atoms, and presents a $T_c$ barely above 1 K~\cite{Cochran58}. Other elemental crystals such as \emph{fcc}-Li, \emph{fcc}-Ir, \emph{bcc}-Nb, and \emph{bcc}-Mo, are included in the analysis in order to establish general trends in metals, where the correlations observed in hydrides do not prevail, as they tend to have high networking values and low $T_c$'s (see Table \ref{tab:table_metals}).

The variety of this sample of superconductors should serve to properly examine the similarities and differences between the derivatives of the KS potentials, and eventually capture useful information that was missing in the electronic contribution addressed previously. For this aim, in the following we define some real-space descriptors that will help us better analyze the derivatives of the KS potential.

\subsection{Contribution to electron-phonon coupling} 

The derivatives of the Kohn-Sham potential present in the matrix elements $g^a_{m\K,n\K'}$ of Eq.~\eqref{eq:e-ph-matrix} are taken with respect to the displacement of one of the atoms in one of its coordinates, represented by the vector $\mathbf{u}_a=u_a \hat{\mathbf{u}}_a$, with $u_a$ the magnitude of the displacement and $\hat{\mathbf{u}}_a$ a unitary vector in the given Cartesian direction. This is then repeated for the rest of the atoms and directions, resulting on a total of $3M$ derivatives that contribute to the electron-phonon coupling, as it can be seen in Eqs.~\eqref{eq:e-ph-eliash} and \eqref{eq:e-ph-delta}. Thus, we define a real-space descriptor that sums the contribution of all of those derivatives,
\begin{align}
    D^2 (\R)= \sum_a \left(\frac{\partial v_{KS}(\R)}{\partial u_a}\right)^2_{u_a=0}\, ,
\end{align} 
where the square is taken in order to avoid cancellation of positive and negative terms.  
Similarly to what was done with the ELF in the previous works, it is possible to unveil the real-space topology of $D^2(\R)$ in the materials of interest by analyzing its isosurfaces and critical points.

\begin{figure*}
    \centering
    \includegraphics[width=0.6\linewidth]{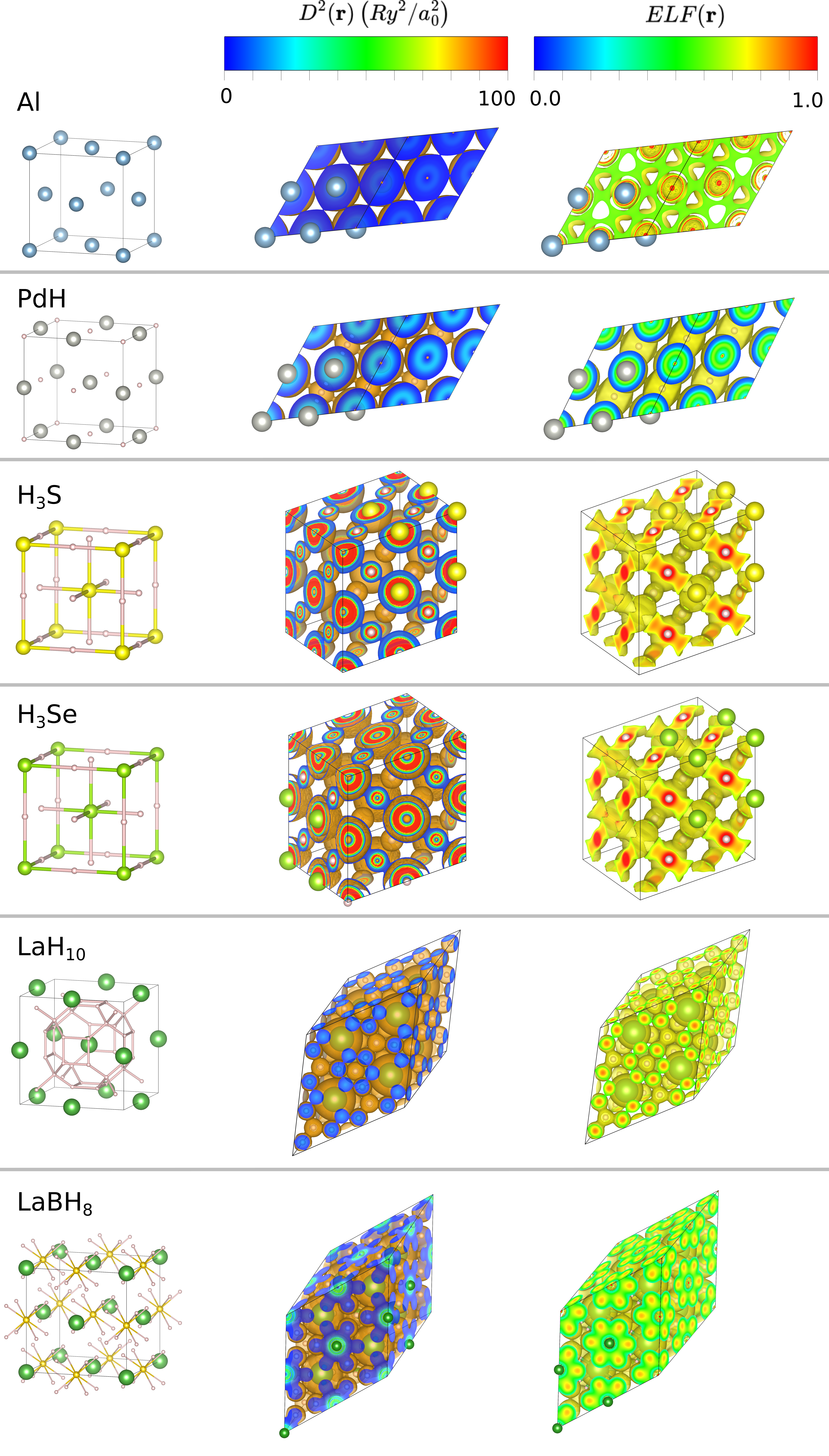}
    \caption{In the left panel, structures of Al, PdH, H$_3$S, H$_3$Se, and LaH$_{10}$. In the center, isosurfaces of $D^2$ at the critical point where all atoms get connected: Al at $D^2=0.38\, Ry^2/a_0^2$, PdH at $D^2=0.94\, Ry^2/a_0^2$, H$_{3}$S (150 GPa) at $D^2=4.8\, Ry^2/a_0^2$, H$_{3}$Se at $D^2=3.8\, Ry^2/a_0^2$, LaH$_{10}$ at $D^2=2.2\, Ry^2/a_0^2$, and LaBH$_{8}$ at $D^2=1.0\, Ry^2/a_0^2$. To the right, ELF isosurfaces of Al at $0.6$, PdH at $0.21$, H$_3$S at 0.7,, H$_3$Se at 0.65, LaH$_{10}$ at 0.56, and LaBH$_8$ at 0.32. }
    \label{fig:D2_all}
\end{figure*}

In Fig.~\ref{fig:D2_all}, one can see that the maxima of $D^2(\R)$ are located on the atomic sites, which is expected, as the profile of the potential is quite steep in that region and any ionic displacement greatly affect its immediate surroundings. In turn, the intermediate values are found along the bonding axes, where the critical points of $D^2(\R)$ are located (see central panel of Fig.~\ref{fig:D2_all}). This can be seen for H$_3$S and LaH$_{10}$, where those critical points 
are along the H-S bond direction, and between the hydrogen atoms, respectively. In fact, the $D^2$ isosurfaces reveal a pattern of connections between nuclei in a similar way than the ELF did
, showing that the types of chemical bonds present in the materials also affect the electron-phonon coupling through the derivatives. This explains the success of the classification of systems into families according to the dominant type of bond appearing in the lattice~\cite{Belli21}. 

In the case of PdH, the topology of $D^2$ is again reminiscent of that of the ELF, as the critical points appear at very low isovalues ($0.94 Ry^2/a_0^2$ compared to $4.8 Ry^2/a_0^2$ for H$_3$S), showing the isolated nature of the bonds. Those lower values anticipate how in this system the contribution of the derivatives to the electron-phonon coupling should be lower, too. 

In the aluminum crystal, the derivatives are low everywhere, 
and the effect of the vibrations on the electronic potential seems to be quite local, as proven by the negligeable values of $D^2(\R)$ in interstitial regions. In contrast, as it can be seen in Fig.~\ref{fig:D2_all}(b), the ELF isosurface shows very delocalized electrons, approaching that of a free electron gas (ELF=0.5). 
This seems to be characteristic of elemental superconductors, with similar features appearing in Li, Ir, Nb, and Mo (see Appendix \ref{sec:metals}).

It seems possible to characterize the metallic bond of those systems by the mismatch between $D^2(\R)$ and the ELF. We observe that, when electron pairs reside in regions where the derivatives of the potential are low (simple metals), $T_c$ is very low (see Table \ref{tab:table_metals}). Meanwhile, when the derivatives are high in those regions, then the transition temperature is enhanced (see H$_3$S, H$_3$Se, LaH$_{10}$, and LaBH$_8$ in Table \ref{tab:table_all-1}). 



In order to have a single value that measures the magnitude of the derivatives for each compound, we define 
\begin{align}
    D^2 = \frac{1}{\Omega_{v}}\int_{\Omega_{v}} D^2(\R)\, d\R\, ,
\end{align}
where the integral is taken over the valence volume, $\Omega_v$. This volume considers the unit cell without the region of the cores, defined by the core radii presented in Table \ref{tab:my_label}. This is done because $D^2(\R)$ is highly dependent on the pseudopotential, and only considering the valence region leads to more consistent results across different theoretical approaches (see Appendix \ref{sec:pp-core_radii}).
The values of $D^2$ are presented in Tables \ref{tab:table_all-1} and \ref{tab:table_metals} for the hydrides and metallic compounds, respectively.

\begin{table}[htbp]
    \centering
    \caption{Summary of reference values and results for the elemental superconductors: pressure (P), networking value ($\phi$), and reference critical temperature obtained from the literature ($T_c^{\rm ref}$), as well as the work from where those were taken.}
    \label{tab:table_metals}
    \begin{tabular}{cccccc}
    \hline
        & P (GPa) & $\phi$ & $D^2(Ry^2/a_0^2)$ & $T_c^{\rm ref}$ (K) & Ref.\\
        \hline \hline
        Al & 0 & 0.61 & 0.44 & 1.2  & \cite{Cochran58} \\
        Li & 33 & 0.59 & 0.14 & 15 & \cite{racioppi2025} \\
        Ir & 5.8 & 0.33 & 0.39 & 0.14 & \cite{kawamura2020}\\
        Nb & 0 & 0.49 & 0.26 & 9.26 & \cite{kawamura2020}\\
        Mo & 0 & 0.47 & 0.3 & 0.92 & \cite{kawamura2020}\\
        \hline \hline
    \end{tabular}   
\end{table}

\begin{figure}[ht]
    \centering
    \includegraphics[width=\linewidth]{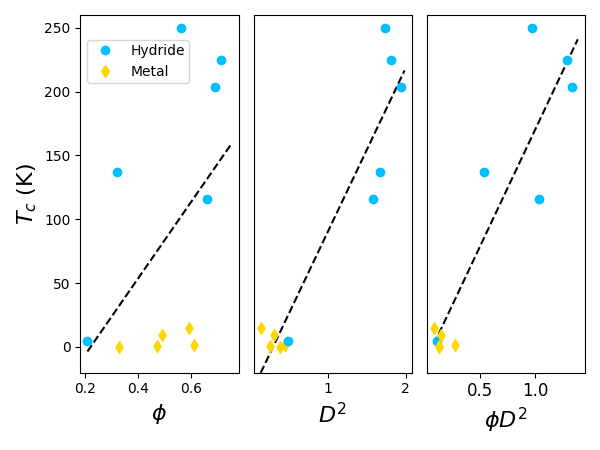}
    \caption{ Critical temperature of conventional superconductors with respect to: networking value, $\phi$ (left), $D^2$ (middle), and the product $\phi D^2$ (right). Dashed lines represent linear fits, and the different markers identify the type of system.} 
    \label{fig:phi-D2}
\end{figure}

In Figure \ref{fig:phi-D2}, we compare the correlation between the critical temperature and the networking value, $\phi$, with that between $T_c$ and the newly defined $D^2$ 
. We observe that the networking value is a very poor predictor for the metallic systems, and that the derivatives of the KS potential, through $D^2$, are able to tell apart low-$T_c$ from high-$T_c$ superconductors. Interestingly, simply combining both quantities does not improve the description.

Additionally, $D^2$ properly measures the 
more pronounced electron-phonon coupling that leads to a higher $T_c$ in H$_3$S with respect to H$_3$Se
, seizing what previous electronic-only-based $T_c$ descriptors failed to capture (see Table \ref{tab:table_all-1}).

\begin{figure}
    \centering
    \includegraphics[width=\linewidth]{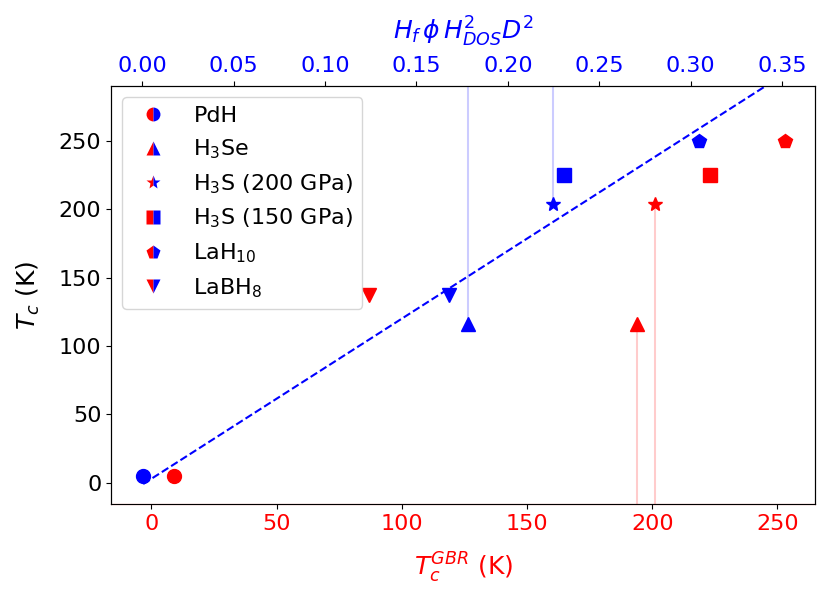}
    \caption{Comparison of the correlations between the reference critical temperature ($T_c$) of hydrides, with two different quantities. In red, 
    with respect to the predictions made with electronic descriptors, $T_c^{GBR}$. In blue, 
    against a new combination of descriptors including derivatives of the KS potential, $H_f\phi H_{DOS}^2 D^2$, with the dashed line representing a linear fit to predict the critical temperature from that quantity. Vertical lines highlight the ability of the second model to distinguish the $T_c$'s of H$_3$Se and H$_3$S at 200 GPa. }
    \label{fig:new-fit}
\end{figure}

Although better at distinguishing those two compounds, $D^2$ does not place the critical temperatures of the hydrides in a perfect growing order, especially in the very high $T_c$ region.
This can be compensated by the other descriptors that have been defined for hydrides, i.e. $\phi$, $H_{DOS}$, and $H_f$. For example, as shown in Fig. \ref{fig:new-fit}, by simply combining those with $D^2$ through the product $H_f\phi H_{DOS}^2 D^2$, 
one recovers a better distribution of $T_c$'s, with LaH$_{10}$ at the top, and properly distinguishing H$_3$S from H$_3$Se. Like this, the correlation with the reference $T_c$ is improved in comparison to what was obtained when predicting only with electronic descriptors, using $T_c^{GBR}$, especially for H$_3$Se. 
Note that the quantity $H_f\phi H_{DOS}^2 D^2$, and the linear fit obtained from its correlation with $T_c$, is only intended to illustrate the power of $D^2$ to improve the characterization of high-$T_c$ compounds, and it is not proposed as a new fit to estimate $T_c$, as the database is too small to justify a generalization. Nonetheless, the integral of $D^2(\R)$ is a good candidate to add to the list of simple descriptors that can be used to fit the critical temperature, e.g. using machine learning. In order to have an improved model that does more accurate predictions, a larger database of hydrogen compounds would be needed. 

\subsection{Beyond the linear approximation} 
\label{sec:non-linear}

The standard approach to assess the magnitude of electron-phonon coupling, both in superconductors as in other materials, is to estimate the matrix elements in Eq.~\eqref{eq:e-ph-matrix} that contain the first derivative of the KS potential. This formulation is actually derived from a Taylor expansion of the potential when the system undergoes a perturbation from its equilibrium ionic positions,
\begin{align}
    \notag
    v_{KS} (\R; \mathbf{R}^0 + \mathbf{u}) =& v_{KS} (\R; \mathbf{R}^0) +\sum_a \left. \frac{\partial v_{KS}}{\partial u_a}\right\vert_{\mathbf{u}=0} u_a  \\ 
    &+ \frac{1}{2} \sum_{a,b} \left.\frac{\partial^2 v_{KS}}{\partial u_a \partial u_b} \right\vert_{\mathbf{u}=0} u_a u_b + \cdots
    \label{eq:taylor_KS}
\end{align}
Then, the appearance of the first derivatives in the matrix elements stems from a linear approximation of the potential,
\begin{align}
    v_{KS} (\R; \mathbf{R}^0 + \mathbf{u}) - v_{KS} (\R; \mathbf{R}^0) \approx \sum_a \left. \frac{\partial v_{KS}}{\partial u_a}\right\vert_{\mathbf{u}=0} u_a \, .
    \label{eq:vks_diff}
\end{align}

To our knowledge, not much has been said about the validity and scope of the approximation made in Eq.~\eqref{eq:vks_diff}. However, there are some reasons to believe that it might be insufficient in hydrides. When one studies the phonon modes of those materials, a similar expansion is done on the Born-Oppenheimer potential upon lattice distortions. In hydrogen-based compounds, the contribution of higher-order derivatives has proven to be significant, and only taking lower-order derivatives leads to important errors in the determination of the phonon modes and the equilibrium geometry. Because in both cases the potentials are expanded with respect to ionic displacements, which can be large for a light atom like hydrogen, a similar result is expected for the KS potential.

In Ref.~\citenum{bianco2023}, electron-phonon coupling was studied from a non-perturbative approach. There, it was shown how a non-linear description of the coupling in palladium hydride, i.e.~taking derivatives beyond first order, greatly affects the estimation of the electron-phonon coupling constant, $\lambda$, while the contrary is observed for a simple metal such as aluminum. It is thus interesting to find a simple way to study the deviations from linearity of the KS potential in different hydrides, and determine the utility of the linear approximation in those systems.


One way to include the changes of $v_{KS}$ in a non-perturbative way is to take its average over an ensemble of distorted configurations. We consider a Gaussian probabilistic ionic distribution $\varrho(\RR)$ centered at $\mathbf{\mathcal{R}}$,
\begin{align}
    \notag
    \varrho(\RR) = &\frac{1}{\sqrt{\det(2\pi\mathbf{\Psi})}} \times \\ 
    & \times \exp\left[-\frac{1}{2} \sum_{a,b} (R_a -\mathcal{R}_a)\Psi_{ab}^{-1}(R_b -\mathcal{R}_b)\right]\, ,
    \label{eq:rho-sscha}
\end{align}
where $\mathbf{\Psi}$ is the displacement-displacement correlation matrix,
\begin{align}
    \notag
    \Psi_{ab} = \left\langle u_a u_b\right\rangle = &\frac{1}{\sqrt{M_a M_b}} \times \\ 
    & \times \sum_{\mathbf{q}\nu} \frac{\hbar (2n_{\mathbf{q}\nu}+1)}{2\omega_{\mathbf{q}\nu}}\, \eta_a(\mathbf{q}\nu)\eta_b({\mathbf{q}\nu})\, ,
    \label{eq:corr-u-u}
\end{align}
with $\eta(\mathbf{q}\nu)$ the polarization vectors, $\omega_{\mathbf{q}\nu}$ the phonon frequencies, $n_{\mathbf{q}\nu}$ the Bose-Einstein occupation numbers and $M_a, M_b$ the mass of the nuclei. In practice, we will take the distribution either from a harmonic approximation, or from the SSCHA, an approach that includes anharmonic effects while using an effective harmonic Hamiltonian (see Appendix \ref{ssec:comput})\cite{bianco2023}.

Using the probability distribution in Eq.~\eqref{eq:rho-sscha}, we take the square of the two sides of Eq.~\eqref{eq:vks_diff}, and average over an ensemble of perturbed configurations,
\begin{align}
    \notag
    \int &\left[v_{KS} (\R; \mathbf{R}^0 + \mathbf{u}) - v_{KS} (\R; \mathbf{R}^0)\right]^2 \varrho(\mathbf{u})\, d\mathbf{u} = \\
    &= \int \sum_{a,b} \left. \frac{\partial v_{KS}}{\partial u_a}\right\vert_{u_a=0}\left. \frac{\partial v_{KS}}{\partial u_b}\right\vert_{u_b=0} u_a u_b \, \varrho(\mathbf{u})\, d\mathbf{u}\, , \\
    &= \sum_a \left( \frac{\partial v_{KS}}{\partial u_a}\right)^2_{u_a=0} \left\langle{u_a^2}\right\rangle_{\varrho} \, .
    \label{eq:vks_diff_sq}
\end{align}
Here, $\left\langle u_a^2 \right\rangle_{\varrho}$ is the mean squared displacement, and it can be obtained from taking the diagonal of the displacement-displacement correlation matrix, 
$\Psi_{aa}$. Both sides of Eq.~\eqref{eq:vks_diff_sq} are real-space functions that can be computed in the compounds studied in this work. For a perfectly linear system, both sides of the equation will be equal. We expect them to evidence larger differences in systems with a higher degree of non-linearity in the electron-phonon coupling.

The right hand side of Eq.~\eqref{eq:vks_diff_sq},
\begin{align}
    (D^2u^2)(\R) &\equiv \sum_a \left( \frac{\partial v_{KS}(\R)}{\partial u_a}\right)^2_{u_a=0} \left\langle{u_a^2}\right\rangle_{\varrho}\, ,
    \label{eq:D2u2}
\end{align}
is very similar to $D^2(\R)$ and can be obtained in an analogous way. The only difference is that the derivatives are weighted by the mean squared displacements of the atoms. Computing the left hand side of \eqref{eq:vks_diff_sq} requires a different approach, and in practice the integral is approximated by an explicit average over a finite number of $N_c$ distorted configurations (see Appendix \ref{ssec:comput}),
\begin{align}
    \notag
    \sigma_{KS}^2 (\R) &\equiv \int \left[v_{KS} (\R; \mathbf{R}^0 + \mathbf{u}) - v_{KS} (\R; \mathbf{R}^0)\right]^2 \varrho(\mathbf{u})\, d\mathbf{u} \, ,\\
    &\approx \frac{1}{N_c} \sum_{I=1}^{N_c} \left[v_{KS} (\mathbf{R}^0 + \mathbf{u}_I) - v_{KS} (\mathbf{R}^0)\right]^2 \, ,
    \label{eq:sigma2}
\end{align}
where $\mathbf{u}_I$ is the displacement from equilibrium of the $I$-th perturbed configuration, obtained according to the probability distribution, $\varrho(\mathbf{u})$. The approximation becomes exact at the limit $N_c\to \infty$.

Computing $(D^2u^2)(\R)$ and $\sigma_{KS}^2(\R)$ on the same compounds as before, we notice that, at first glance, the real-space isosurfaces of Eqs.~\eqref{eq:D2u2} and \eqref{eq:sigma2} do not provide much information about the magnitude of the non-linear terms in the systems of study (see Appendix \ref{sec:isosurfaces}). Thus, we again integrate the valence regions of the supercell to define the global quantities,
\begin{align}
    D^2u^2 &= \frac{1}{\Omega_v} \int_{\Omega_v} (D^2u^2)(\R)\, d\R\, ,\\
    \sigma_{KS}^2 &= \frac{1}{\Omega_v} \int_{\Omega_v} \sigma_{KS}^2(\R)\, d\R\, .
\end{align}

As it can be seen in Table \ref{tab:table_all2}, the results show that in the simple metals the differences between $\sigma^2$ and $D^2u^2$ are generally smaller, and therefore the  linear approximation will be more adequate for those systems. In hydrogen-based superconductors, on the other hand, the differences are at least an order of magnitude larger than in most metals, making such an approximation less adequate.


\begin{table}[ht!]
    \centering
    \setlength{\tabcolsep}{0.7\tabcolsep}
    \caption{Values of descriptors that measure the validity of the linear approximation in hydrides and Al, including the absolute difference between $\sigma_{KS}^2$ and $D^2 u^2$, and $\Delta V_{KS}$.}   
    \label{tab:table_all2}
    \begin{tabular}{ccc}
    \hline
        & $\abs{\sigma_{KS}^2- D^2u^2} \left(Ry^2\right)$   & $\Delta V_{KS} (Ry)$ \\
        \hline \hline
        Al &  $2.4\times 10^{-4}$ & $3.4\times 10^{-3}$ \\ 
        Li & $1.6\times 10^{-3}$ & $3.2\times 10^{-3}$\\ 
        Ir & $5.4\times 10^{-6}$ & $1.0\times 10^{-3}$\\ 
        Nb & $1.4\times 10^{-4}$ & $1.4\times 10^{-3}$\\ 
        Mo & $8.9\times 10^{-5}$ & $1.6\times 10^{-3}$\\ 
        PdH & $1.4\times 10^{-2}$ & $1.4 \times 10^{-2}$\\ 
        H$_3$Se & $6.6\times 10^{-3}$ & $2.2\times 10^{-2}$\\ 
        H$_3$S (200 GPa) & $5.1\times 10^{-3}$ & $2.4\times 10^{-2}$ \\
        H$_3$S (150 GPa) & $5.3 \times 10^{-3}$ & $2.3\times 10^{-2}$\\ 
        LaH$_{10}$ & $7.7\times 10^{-3}$ & $2.7\times 10^{-2}$\\ 
        LaBH$_{8}$ & $5.6\times 10^{-3}$ & $2.2\times 10^{-2}$\\ 
        \hline \hline
    \end{tabular}
\end{table}

Another way of assessing non-linearity can be obtained if one does not take the square before integrating in Eq.~\eqref{eq:vks_diff_sq}, obtaining a quantity that is formally zero, due to the parity of the gaussian probability distribution,
\begin{align}
    \label{eq:diff_vks_int1}
    \Delta v_{KS}(\R) &\equiv \int \left[v_{KS} (\R; \mathbf{R}^0 + \mathbf{u}) - v_{KS} (\R; \mathbf{R}^0)\right] \varrho(\mathbf{u})\, d\mathbf{u} \\
    &= \int \sum_a \left. \frac{\partial v_{KS}}{\partial u_a}\right\vert_{u=0} u_a \varrho(\mathbf{u})\, d\mathbf{u} = 0\, .
    \label{eq:diff_vks_int2}
\end{align}
Then, the global descriptor
\begin{align}
    \Delta V_{KS} =  \frac{1}{\Omega_v} \int_{\Omega_v}  \abs{\Delta v_{KS}(\R)}\, d\R\, ,
\end{align}
will only be different from zero when there is a deviation from the linear regime. Here, the absolute value was considered in order to avoid cancelation of positive and negative terms. The results are collected in Table \ref{tab:table_all2}.

The non-linear effects measured by $\Delta V_{KS}$ are easier to visualize, and again we see that the systems containing hydrogen show deviations from linearity an order of magnitude larger than that of aluminum. In particular, the results obtained for aluminum and palladium hydride are in agreement with previous observations, as the former is known to have a linear behavior, whereas in the latter this approximation has proved to be insufficient \cite{bianco2023}. We are not aware of any previous studies on the importance of the non-linear terms of the KS potential for other hydrides. 
The descriptors proposed in this work predict that indeed the non-linear effects will be important across all the hydrides here presented. This will impact important quantities depending on electron-phonon matrix elements such as the Eliashberg function, coupling constant, and critical temperature. It thus becomes essential to continue to develop new methods that allow the inclusion of non-linear effects and more accurately represent the underlying physics.


\section{Conclusions}
\label{sec:conclusions}

In this work, we studied the changes of the KS potential under ionic perturbations from equilibrium, and analyzed their effect on the electron-phonon coupling. This was done with two different aims: to assess the importance of the spatial resolution of the derivatives of $v_{KS}$ compared to the electron localization function
; and to understand the limitations of approximating the KS potential linearly with respect to those changes. In both cases, real-space local and global quantities were proposed and later evaluated on well-known superconductors, considering compounds with and without hydrogen.

The spatial representation of the derivatives of $v_{KS}$ obtained with $D^2(\R)$ in hydrogen-based compounds showcased a qualitative correlation between the regions of higher variations of the potential and those of higher electron localization, as determined by the ELF. This analysis further supports the classification of materials according to their bonding type first presented in Ref.~\citenum{Belli21}~, which serves as a tool to recognize promising structures with higher critical temperatures. The correlation between those spatial patterns 
explains the success of the descriptors stemming from the ELF to predict the critical temperatures. Nonetheless, the information contained in the derivatives of the potential is necessary to construct a more complete picture of electron-phonon coupling. In fact, in the metallic compounds, both functions display completely different shapes and, while the high electron delocalization is a signature of the metallic character of the system, the fluctuations of the KS potential far from the cores are not as important, which results in weak coupling. Indeed, a problem with the ELF-derived descriptors is that it could create \emph{false positives} when there is a good metal, which is not necessarily conducive to a high $T_c$. In contrast, $D^2$ consistently shows a reliable characterization, providing new information that was lacking in the electronic description, and properly predicting their low $T_c$. 

In the case of hydrides, $D^2$ successfully captures the difference in the electron-phonon coupling of systems with similar bonding patterns, such as H$_3$S and H$_3$Se in the $Im\bar{3}m$ phase. 
The discrepancy between their critical temperatures is explained by the different magnitudes of the derivatives of $v_{KS}$, as measured by $D^2$. The proposed descriptor is thus a good candidate to join other quantities used in the fast first estimations of $T_c$ in superhydrides.

The assessment of the fitness of the linear approximation in the study of electron-phonon coupling showed that the approach is insufficient to fully describe the phenomenon in hydride superconductors. This demonstrates the non-linearity of the potential in those materials, as it had been previously discussed for the case of palladium hydride \cite{bianco2023}. Our method predicts H$_3$S, H$_3$Se, LaH$_{10}$, and LaBH$_8$ to be non-linear, too, meaning that care must be taken when $\lambda$ is computed in the standard (linear) approach, as it might not be as exact as expected. Ideally, as it is done in Ref.~\citenum{bianco2023} for Al and PdH, higher order derivatives should be performed to obtain more accurate results. Alternatively, the simplicity of the statistically averaged descriptors proposed in this contribution can serve as a practical tool to diagnose the degree of non-linearity of the potential in the system. 

\section*{Acknowledgments}
This work is supported by the European Research Council (ERC) under the European Unions Horizon 2020 research and innovation program (Grant Agreement No. 802533 and 810367), the Spanish Ministry of Science and Innovation (Grant No. PID2022142861NA-I00), the Department of Education, Universities and Research of the Eusko Jaurlaritza and the University of the Basque Country UPV/EHU (Grant No. IT1527-22), IKUR strategy--an initiative of Department of Science, Universities and Innovation of the Basque Government, and Simons Foundation through the Collaboration on New Frontiers in Superconductivity (Grant No. SFI-MPS-NFS-00006741-10). The authors also thank ANR TcPredictor S22-ERC036, ANR Fiscency S23-JRAR060, EMERGENCE-SU H20X S23JR31014, GENCI A0040710377 for funding and resources.

\appendix

\section{Methods}
\label{ssec:comput}

The Kohn-Sham potentials used throughout this work, as well as the ELF and the DOS, are obtained during post-processing of a self-consistent field (SCF) calculation performed using the Quantum ESPRESSO plane-wave package.\cite{QE09,QE17} Only the local component of the KS potential is considered in the analysis. The estimated critical temperature, $T_c^{GBR}$, was obtained using TcESTIME.\cite{novoa2024} For each material, a 2x2x2 supercell was used in order to properly describe the effect of phonon fluctuations. The lattice parameters and atomic positions of PdH, H$_3$S at 150 GPa, LaH$_{10}$, and LaBH$_8$ have been taken from the literature \cite{errea2013,errea2016,errea2020,belli2022}. For H$_3$Se and H$_3$S at 200 GPa, the initial structure was taken from the literature, \cite{zhang2015,duan2014} but were later relaxed, obtaining the lattice constants $a=3.153$~\AA~and $a=2.988$~\AA~for the unit cell, respectively. The fcc Al, Li and Ir crystals were also relaxed, yielding unit cell lattice constants of $4.042$~\AA, $3.403$~\AA, and $3.850$~\AA, respectively; while for the bcc Nb and Mo lattices the values of $3.242$~\AA~ and $3.140$~\AA~ were obtained. DFT calculations are done within generalized-gradient approximation, using the PBE exchange-correlation functional \cite{PBE}. Plane-augmented wave (PAW) pseudopotentials are employed for all the elements, except Nb and Mo, for which a norm-conserving pseudopotential without semicore electrons was employed (see Appendix \ref{sec:pp-core_radii}). For each system, a Methfessel–Paxton smearing of 0.02 Ry was used, whereas the energy cutoff for the wavefunctions and density were converged. Those were set to 40 and 400 Ry, respectively, for Al. For LaH$_{10}$, they were set to 50 and 500 Ry, to 60 Ry and 600 Ry for H$_3$Se and Ir, to 70 Ry and 700 Ry for H$_3$S, PdH, Nb, and Mo, and to 90 and 900 for Li and LaBH$_8$.
The k-points were sampled using a uniform Monkhorst-Pack grid, which was also converged for each compound. A grid of size 6x6x6 was used for LaH$_{10}$, of 10x10x10 for PdH and LaBH$_8$, of 16x16x16 for Li, Nb and H$_3$S at 150 GPa, of 18x18x18 for Al, 20x20x20 for Ir, 22x22x22 for H$_3$S at 200 GPa, 24x24x24 for Mo, and 26x26x26 for H$_3$Se.

Phonon calculations were performed on all systems, in order to obtain the dynamical matrices needed in the computation of the ionic probability distribution, $\varrho(\mathbf{u})$, and the root mean squared displacements. In the case of PdH, LaH$_{10}$, LaBH$_8$, and H$_3$S at 150 GPa, this was done at the stochastic self-consistent harmonic approximation (SSCHA) level, with the details of the calculations being the same as in Refs.~\cite{errea2013,errea2020,belli2022,errea2016}. For the other compounds, 
the harmonic approximation was employed, using density functional perturbation theory as implemented in Quantum ESPRESSO. 

The derivatives of the KS potential are obtained using finite differences,
\begin{align}
    \left.\frac{\partial v_{KS}(\R; \RR^0+u_a)}{\partial u_a} \right\vert_{u_a=0} \approx \frac{v_{KS}(\R; \RR^0+\mathbf{u}_{a})-v_{KS}(\R; \RR^0)}{u_a} \, ,
\end{align}
taking $\mathbf{u}_{a} =u_a \hat{\mathbf{u}}_a$ as a vector of length $u_a=0.01$\AA\, displacing the $i$-th ion in the cartesian direction $\alpha$, where $a=(i,\alpha)$.  Thus, the total amount of SCF calculations per system are 6M, with M the number of atoms in the supercell. Depending on $M$, computation times can range between 20 CPU minutes to a few CPU hours ($\sim 5$ CPU hours for LaH$_{10}$, the largest system). The independence of the SCF calculations ensures efficient parallelization, reducing this to wall-time minutes.

The resulting derivatives are summed to get $D^2(\R)$. A similar thing is done for $D^2u^2(\R)$, but in that case the root mean squared displacement is also needed, $\left\langle u_a^2 \right\rangle$. It is obtained through Eq.~\eqref{eq:corr-u-u} and computed using the CellConstructor package of the SSCHA Python-API.\cite{monacelli2021}

As it was previously mentioned, the statistical averages needed to compute $\sigma_{KS}^2(\R)$ and $\Delta v_{KS}(\R)$ are done in an approximate fashion, using a number of $N_c$ perturbed configurations (see Eq.~\eqref{eq:sigma2}).
The distorted positions are obtained from the probability distribution $\varrho(\mathbf{u})$, also computed with the CellConstructor package. To accelerate convergence with respect to $N_c$, we increase the number of configurations by using the symmetry operations of the crystal. Then, if $\mathbf{S}$ is 
one of those symmetry operations, and $v_{KS}(\R, \RR_I)$ is known, with $\RR_I = \RR^0+\mathbf{u}_I$, we can rotate the grid of points $\R$ to obtain
\begin{align}
v_{KS} (\R; \mathbf{S}\RR_I) = v_{KS} (\mathbf{S}^{-1}\R; \RR_I)\, ,
\label{eq:symm}
\end{align}
In this way, we have access to the KS potential for a perturbed configuration with ionic positions $\mathbf{S}\RR_I$, without the need of doing a SCF calculation. Like this, we can include in Eq.~\eqref{eq:sigma2} as many perturbed configurations per DFT calculation as the number of symmetry operations of the crystal, $N_{symm}$. For each system, $50$ distinct distorted configurations, leading to a total of $N_c=50N_{symm}$ terms in the sum in the estimation of $\sigma_{KS}^2(\R)$ and $\Delta v_{KS}(\R)$, were considered.

\section{Pseudopotentials and core radii}
\label{sec:pp-core_radii}

\begin{figure}[ht!]
    \centering
    \includegraphics[width=0.8\linewidth]{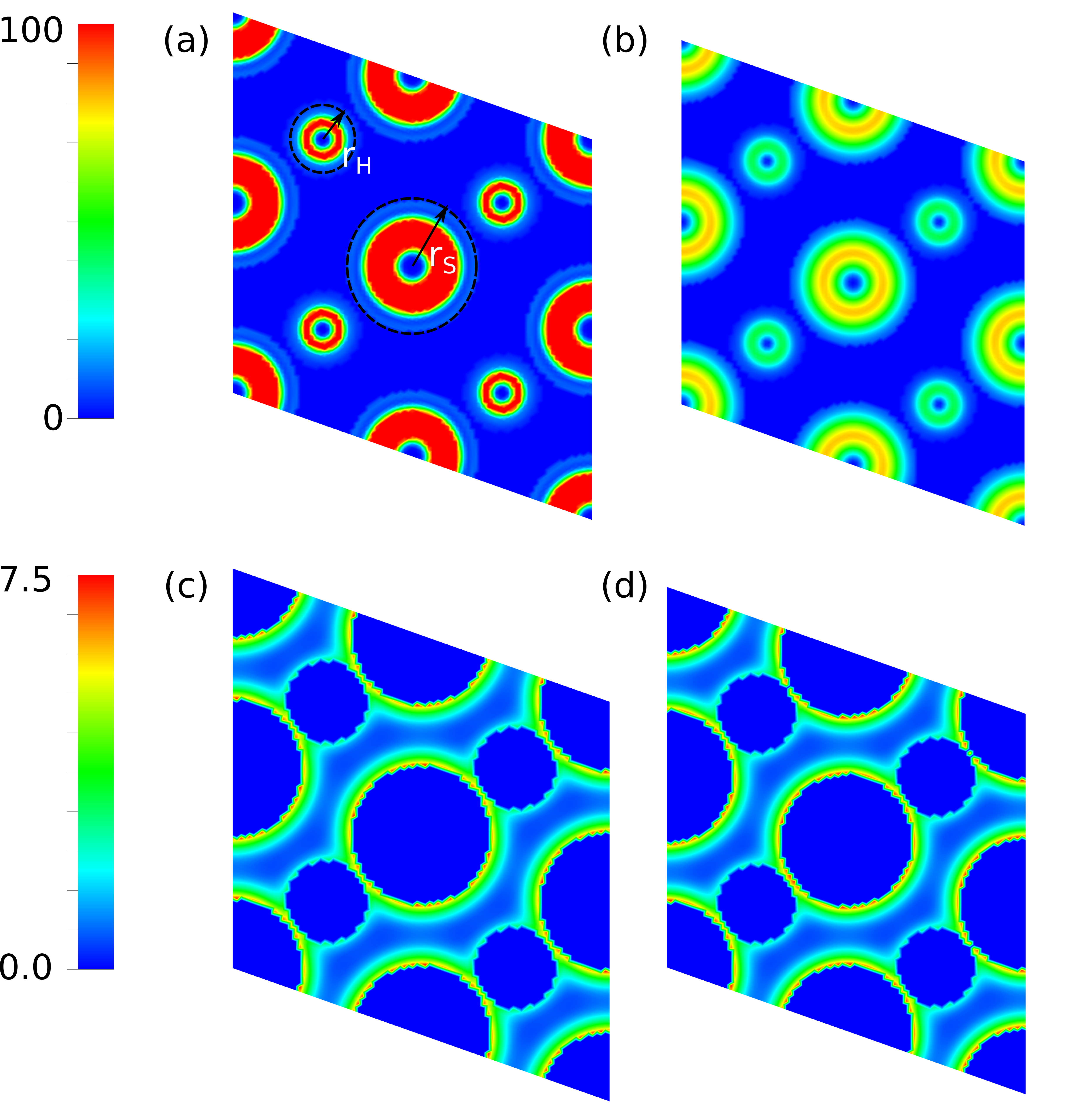}
    \caption{Comparison of $D^2(\R)$ values along a plane in the H$_3$S crystal at 200 GPa, inside and outside the cores, defined by the radius $r_S=1.7\, a_0$ for sulfur and $r_H=1.0\, a_0$ for hydrogen, as depicted in (a). On the top panel, the colormap shows the values inside the core when using (a) plane augmented wave and (b) norm-conserving pseudopotentials. Using the same pseudopotentials, the values of $D^2(\R)$ outside the cores are depicted in (c) and (d), respectively.}
    \label{fig:core-nocore}
\end{figure}

The pseudopotential changes the observed KS potential and its derivatives inside a cutoff radius $r_c$, as shown in Figure \ref{fig:core-nocore}. Once the core has been removed from the integration the result becomes independent of the pseudopotential used. 

In Table \ref{tab:my_label} we specify the radii used for the integration of the core. In most cases, it corresponds to the maximum radius specified in the pseudopotential file, including the local part and all momentum channels. 

In the case of Nb and Mo, the norm-conserving Hartwigesen-Goedecker-Hutter pseudopotentials with 5 and 6 electrons, respectively, are used. This is because it proves better to exclude semi-core electrons from the calculations of transition metals, as their electronic bands lay far from the Fermi energy, and do not contribute to electron-phonon coupling. The cutoff radii for those pseudos have been obtained by visual inspection of all-electron and pseudo-wavefunctions.

    \begin{table}
    \centering
    \begin{tabular}{ccc}
    \hline
        Element & $r_{c}\,  (a_0)$  \\
    \hline \hline
        H &  1.00 \\
        Li & 2.00  \\
        B  & 1.40 \\
        Al & 2.00 \\
        S &  1.70 \\
        Se & 1.90\\
        Nb & 2.6  \\
        Mo & 2.5  \\
        Pd & 2.60 \\
        La & 2.00 \\
        Ir & 2.70  \\
    \hline\hline
    \end{tabular}
    \caption{Cutoff radius, $r_c$, used in the integration of $D^2$ and other real-space quantities.}
    \label{tab:my_label}
\end{table}

\section{Isosurfaces of $D²(\R)$ in elemental metals}
\label{sec:metals}

\begin{figure*}
    \centering
    \includegraphics[width=0.6\linewidth]{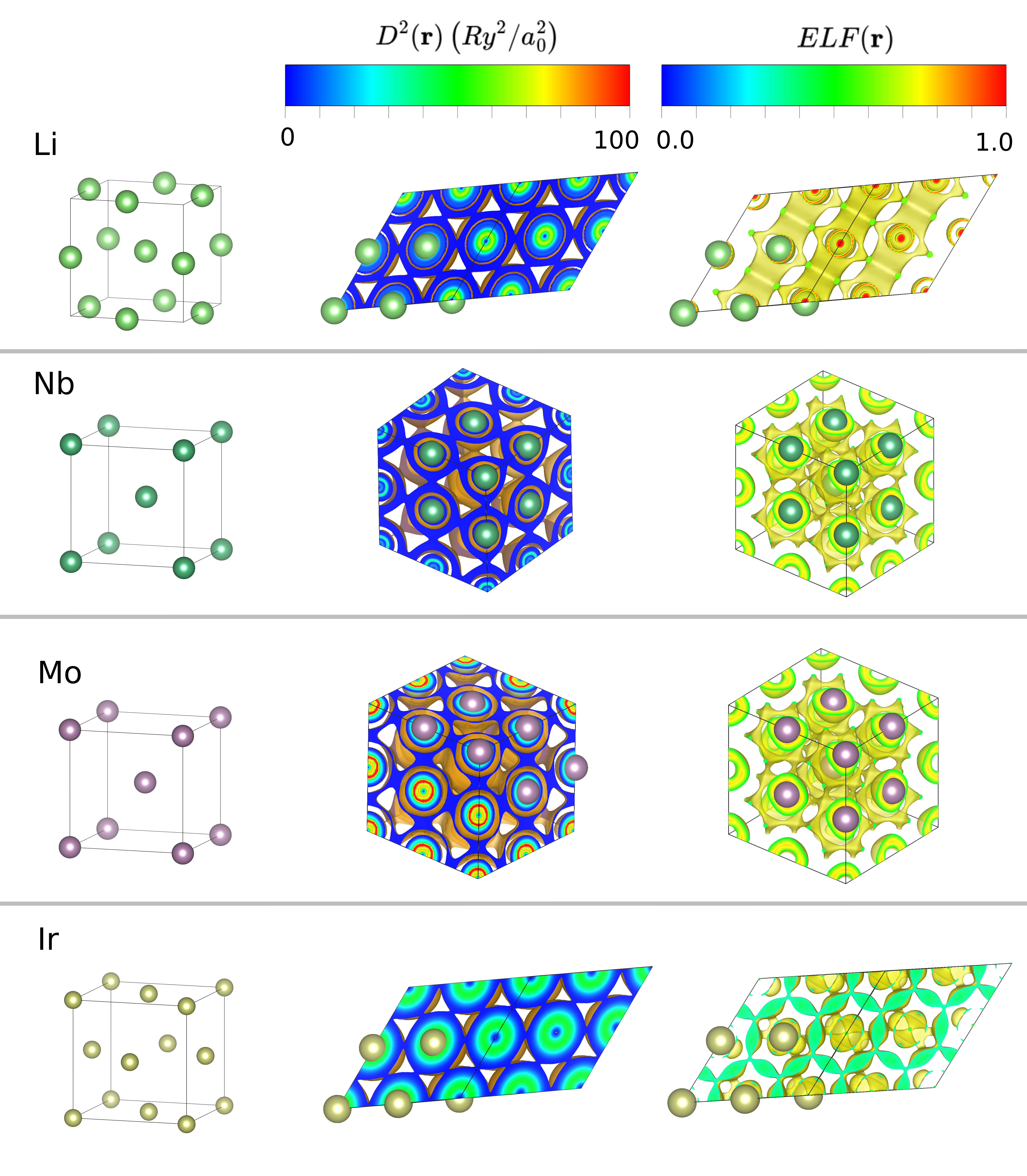}
   \caption{In the left panel, structures of Li, Nb, Mo, and Ir. In the center, isosurfaces of $D^2$ at the critical point where all atoms get connected: Li at $D^2=0.0.18\, Ry^2/a_0^2$, Nb at $D^2=0.55\, Ry^2/a_0^2$, Mo at $D^2=0.48\, Ry^2/a_0^2$, and Ir at $D^2=1.2\, Ry^2/a_0^2$. To the right, ELF isosurfaces of Li at $0.59$, Nb at $0.45$, Mo at $0.43$, and Ir at $0.33$. }
    \label{fig:metals}
\end{figure*}

In Fig. \ref{fig:metals} we show isosurfaces of $D^2(\R)$ in elemental metals.

\section{Isosurfaces of other descriptors}
\label{sec:isosurfaces}

\begin{figure}[ht!]
    \captionsetup[subfigure]{position=t}
    \centering
    \subcaptionbox{Al, $D^2u^2=0.01\, Ry^2$}{%
\includegraphics[width=0.3\linewidth]{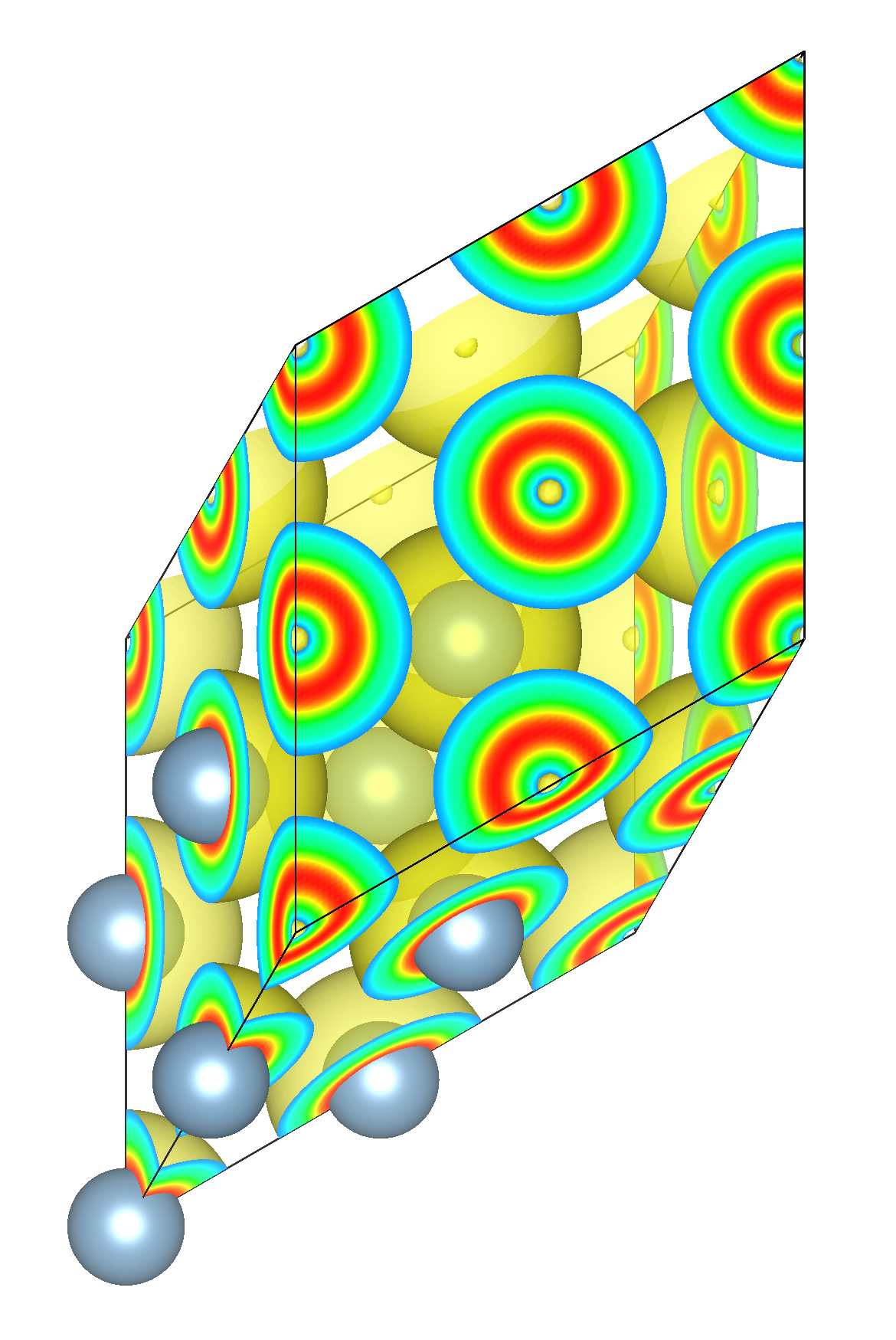}}
\subcaptionbox{Al, $\sigma_{KS}^2=0.01\, Ry^2$}{%
\includegraphics[width=0.31\linewidth]{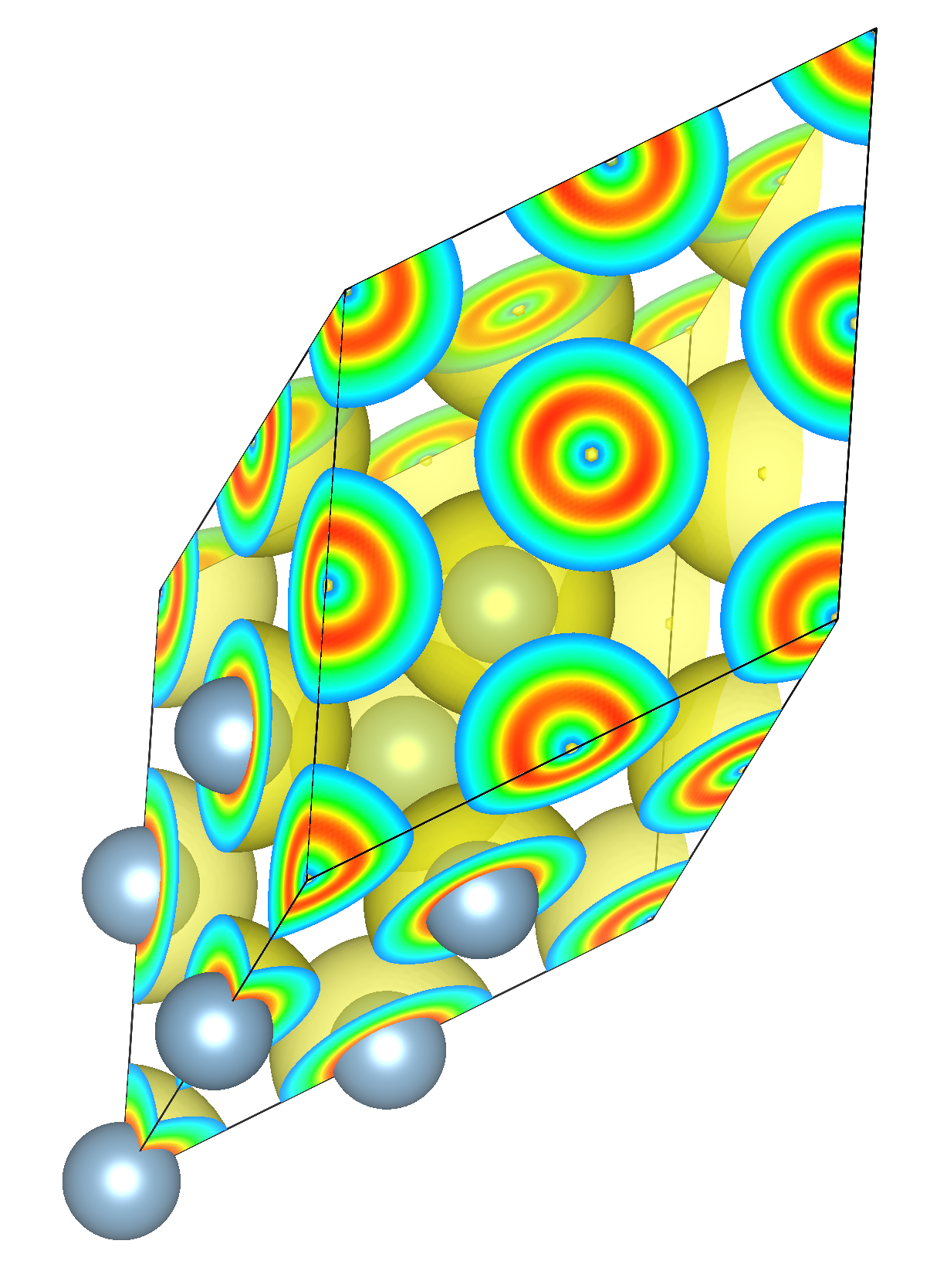}}\\
\subcaptionbox{PdH, $D^2u^2=0.015\, Ry^2$}{%
\includegraphics[width=0.36\linewidth]{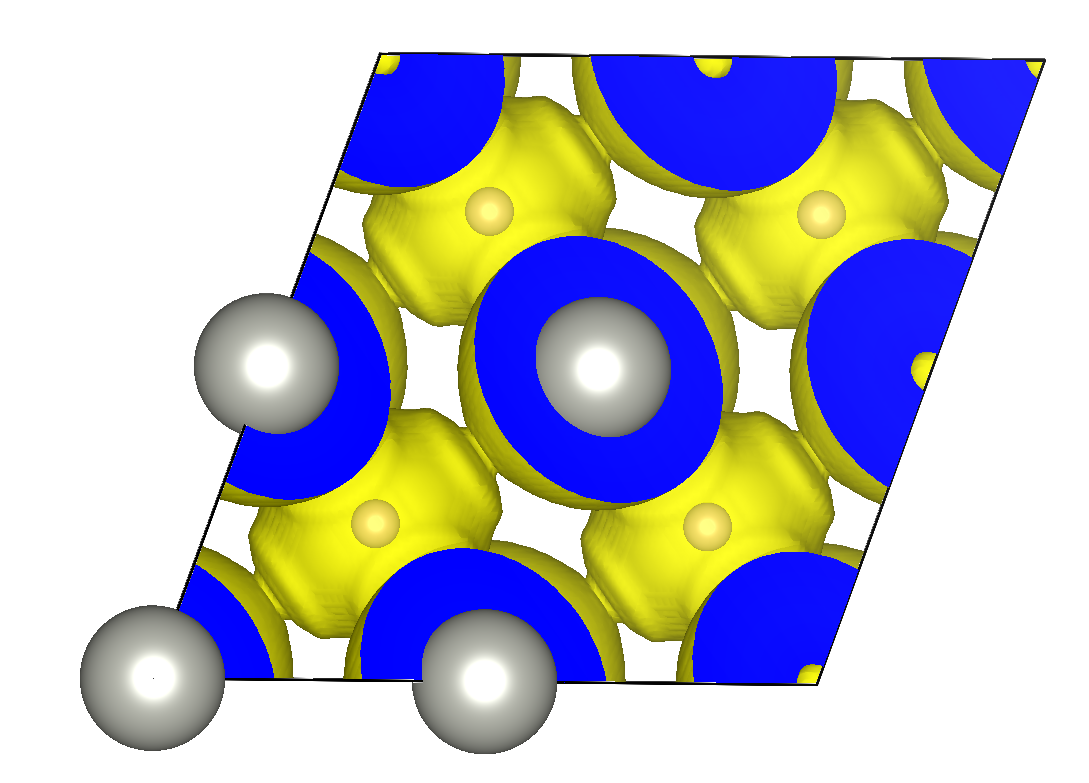}}
\subcaptionbox{PdH, $\sigma_{KS}^2=0.015\, Ry^2$}{%
\includegraphics[width=0.36\linewidth]{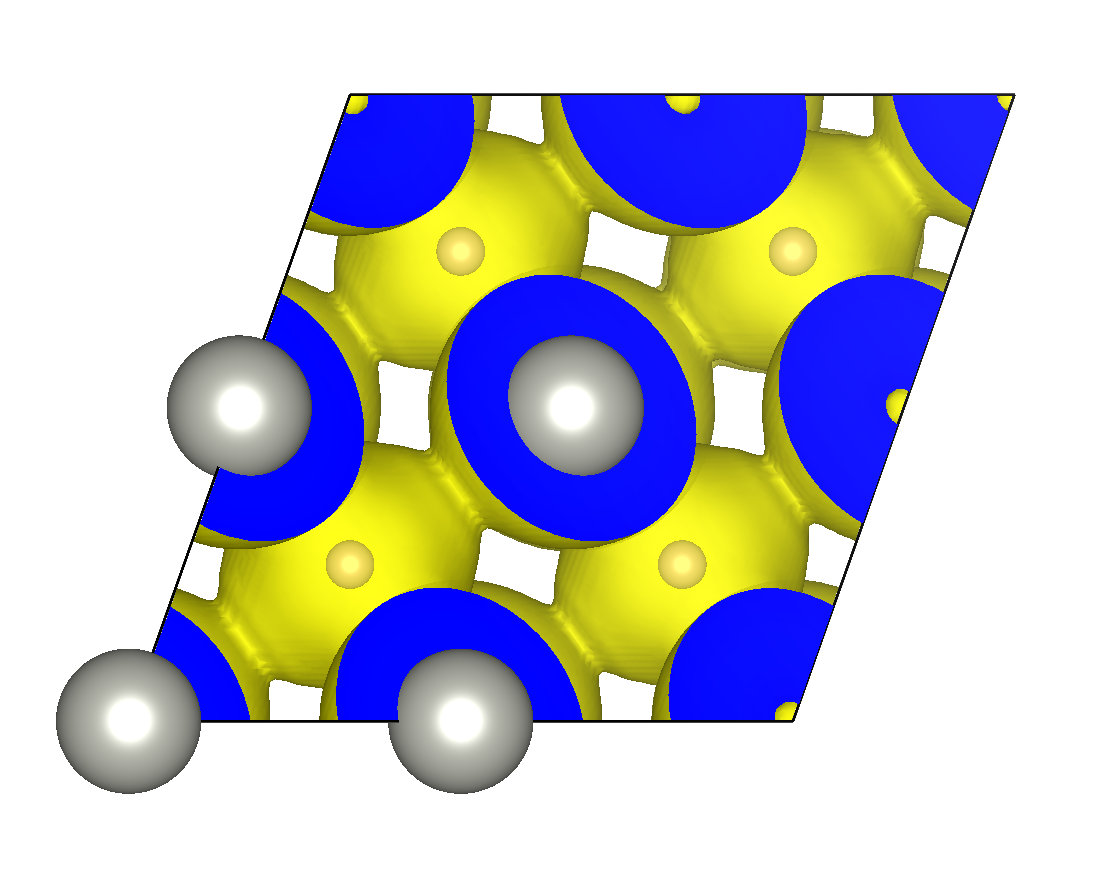}}\\
\subcaptionbox{H$_3$S, $D^2u^2=0.03\, Ry^2$}{%
\includegraphics[width=0.33\linewidth]{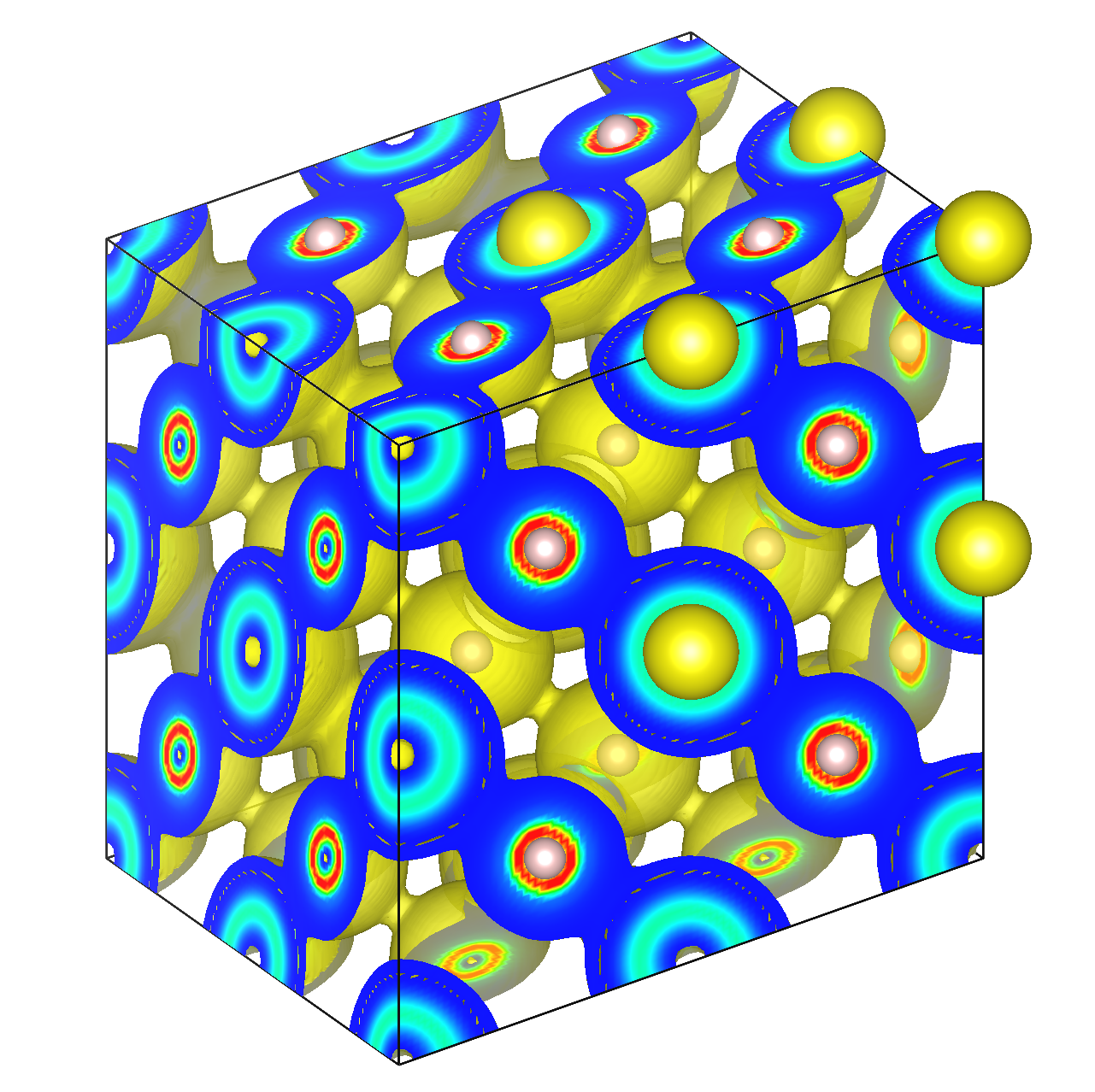}}
\subcaptionbox{H$_3$S, $\sigma_{KS}^2=0.03\, Ry^2$}{%
\includegraphics[width=0.33\linewidth]{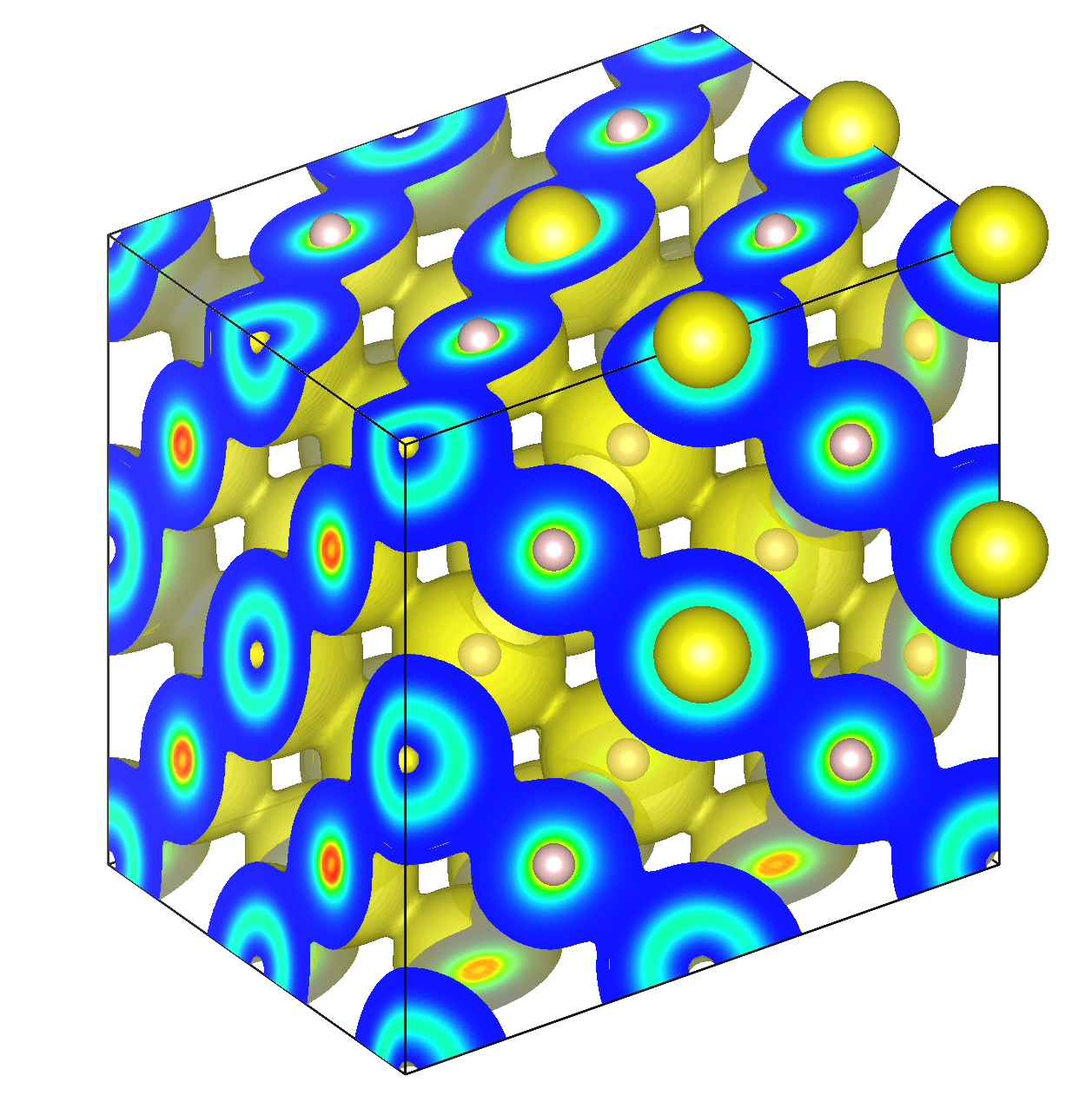}}
    \caption{Comparison between isosurfaces of $D^2u^2(\R)$ and $\sigma_{KS}^2(\R)$ in Al, PdH and H3S at 150 GPa.}
    \label{fig:enter-label}
\end{figure}

In Fig. \ref{fig:enter-label} we show isosurfaces of $D^2u^2(\R)$ and $\sigma_{KS}^2(\R)$ for comparison.

\newpage
\bibliography{supercond}

\end{document}